\documentclass[a4paper,11pt]{article}
\usepackage[T1]{fontenc}
\usepackage{jcappub} 
\usepackage{aas}
\usepackage{lineno}
\usepackage{multirow}
\usepackage{bm}

\arxivnumber{2510.26767} 

\title{Unbiased Primordial Gravitational Wave Inference from the CMB with SMICA}

\author[a]{Alexander Steier,}
\author[a, b]{Shamik Ghosh,}
\author[b, a]{and Jacques Delabrouille,}
\affiliation[a]{Lawrence Berkeley National Laboratory, 1 Cyclotron Road, Berkeley, CA 94720, USA}
\affiliation[b]{CNRS-UCB International Research Laboratory, Centre Pierre Bin\'etruy, IRL 2007, CPB-IN2P3, Berkeley, CA 94720, USA}

\abstract{The detection of primordial gravitational waves in Cosmic Microwave Background B-mode polarization observations requires accurate and robust subtraction of astrophysical contamination. We show, using a blind Spectral Matching Independent Component Analysis, that it is possible to infer unbiased estimates of the primordial B-mode signal from ground-based observations of a small patch of sky even for highly complex foreground contamination. This work, originally performed in the context of configuration studies for a future CMB-S4 observatory, is highly relevant for the analysis of observations by the current generation of CMB experiments.}

\begin{document}
\maketitle
\flushbottom

\section{Introduction}

The Cosmic Microwave Background (CMB), relic radiation that decoupled from the primordial plasma when the electrons and nuclei first combined to form neutral atoms, carries a wealth of information about the physical processes that govern the history and the matter-energy content of the Universe. 
Tiny fluctuations of the spacetime metric in the early Universe leave small imprints on the observed temperature and polarization of the CMB. 
Polarization signals are customarily decomposed into even-parity $E$ modes and odd-parity $B$ modes \cite{ZaldarriagaSeljak1997, KamionkowskyKosowskyStebbins1997b}. 
Scalar perturbations of the metric, corresponding to density inhomogeneities, generate most of the temperature anisotropies of the CMB, as well as most of the $E$-mode polarization, but do not directly generate $B$ modes. 
Tensor perturbations however, corresponding to primordial gravity waves, would generate both $E$ and $B$ modes. 

One of the most compelling theories for the origin of cosmic structures postulates that initial density perturbations, which seed large-scale structures such as galaxies and clusters of galaxies, are generated during an epoch of fast accelerated expansion in the early universe, known as cosmic inflation.
Many of the simplest inflationary scenarios predict the existence, in addition to density perturbations, of a primordial gravitational wave (PWG) background in the form of random tensor perturbations of the metric, which could be detectable in the form of tiny parity-odd $B$-mode patterns in the linear polarization of the CMB~\cite{2016ARA&A..54..227K}. 
Their unambiguous detection would be a compelling confirmation of cosmic inflation, and as various models predict different levels of tensor perturbations, quantified by the ratio of the primordial tensor power to the primordial scalar power ($T/S=r$), a detection would also provide information on which of many possible inflationary scenarios best describes the physics at work at energies well beyond the reach of human-made particle accelerators \cite{2014PDU.....5...75M}. 

Several ongoing \cite{BICEP2021, SO2019, AliCPT2022} and future \cite{LiteBIRD2024, PICO2019} CMB experiments specifically target a first detection of CMB polarization $B$ modes originating from PWGs.
This objective is experimentally challenging, as the expected signal is tiny. 
Reaching the sensitivity necessary for a detection requires accumulating years of observations with large state-of-the-art CMB telescopes, from the ground or from space. 
For example, the CMB Stage-4 (CMB-S4) experiment was designed to deploy hundreds of thousands of cryogenic detectors in the focal planes of a collection of refractive telescopes located at the South Pole in Antarctica, with the aim of reaching a sensitivity of $\sigma_r < 5\times 10^{-4}$ on the tensor-to-scalar ratio \cite{CMBS42022}. 
Besides instrumental noise, spurious CMB $B$ modes generated by gravitational lensing of original $E$ modes by large-scale structures along the photon paths constitute an additional source of potential confusion, and polarized microwave emission from the interstellar medium of our own Galaxy generates foreground microwave polarization signals that are orders of magnitude brighter than the target PGW signal. 

Since their inception, CMB experiments have addressed the challenge of separating foreground emissions from the CMB by combining multi-frequency observations in the 20-900 GHz range \cite{Martinez-Gonzalez2003, Eriksen2004, Delabrouille2009, Eriksen2008, Stompor2009}. 
Some methods make no assumptions about the nature of the foregrounds \cite{Martinez-Gonzalez2003, Eriksen2004, Delabrouille2009}, while others rely on specific parametric models for the emission laws and statistical properties of the foreground signal \cite{Eriksen2008, Stompor2009}. 
Depending on the details of the implementation and the complexity of the foreground emission, all of these methods can produce biased measurements of the tensor-to-scalar ratio $r$ by reason of residual foreground contamination. 
Such residuals can be mitigated by either adding additional terms in the data model, or by marginalizing over a model of residual foregrounds, but those approaches typically involve modeling quantities that are neither well constrained nor well understood at the level of sensitivity required for $r$-measurement.  

The Spectral Matching Independent Component Analysis (SMICA)~\cite{Delabrouille03, Cardoso08} is a method that models the multivariate power spectrum of multi-frequency observations as a sum of contributions from independent sources of emission. 
SMICA is a flexible blind parametric method which, unlike other parametric methods of component separation, can be implemented without making any prior assumptions about the foregrounds. 
In the past, SMICA has been used to produce CMB and foreground maps from WMAP and Planck space mission observations \cite{2005MNRAS.364.1185P,2014A&A...571A..12P,2020A&A...641A...4P}. 
In the present work, we study the performance of a multi-component SMICA model for a CMB-S4-like experiment on a low-foreground patch of sky in the Southern Galactic hemisphere. 
While this study is tailored for a CMB-S4-like instrumental configuration, the conclusions would translate to Stage-3 CMB experiments such as BICEP-Keck or the Simons Observatory.

This paper is organized as follows: we describe our sky simulation pipeline in section~\ref{sec:mockobservations}, the data model, likelihood, SMICA pipeline implementation in section~\ref{sec:pipeline}, and results for different foreground complexities in section~\ref{sec:results}. 
Those results are discussed in section~\ref{sec:discussion}, and we conclude in section~\ref{sec:conclusions}.

\section{Mock Observations}
\label{sec:mockobservations}

In this work we consider instrument parameters and science targets in line with those of CMB-S4. 
We define a possible sky patch that would be similar to the CMB-S4 `ultra-deep' sky patch \cite{CMBS42022}, with a noise model adapted from the most recently published CMB-S4 forecasting paper \cite{S4pipeline25}, as detailed in Table~\ref{tab:S4specs}. Two options are considered: One of them, called ``non-split'', has a total of 7 frequency bands, with two main CMB channels at 90 and 150~GHz, three low frequency channels, and two high frequency channels. 
The other, labelled as ``split'', replaces the two CMB channels by four channels centered at 85, 95, 145, and 155 GHz, for a total of 9 bands.
The various frequency channels are selected for discerning between CMB signals, the frequency dependence of which are those of the temperature derivative of a 2.726~K blackbody, and those of Galactic foreground emissions. 
We generate 96 different noise realizations, and an equal number of CMB realizations, for a statistically significant sample of mock observations to test the pipeline.

We note that the map-based simulations used in this work do not capture all possible complexities of the data properties of ground-based CMB measurements. Processing of the data timestreams with various filters and template subtractions would produce effects and residual systematics that are not captured by our simulation pipeline. We postpone the inclusion of such effects and systematics to a future project.

\begin{table}[tb]
\centering
\begin{tabular}{c|c|c|c|c|c}
\hline
      Frequency      & $\theta_\text{FWHM}$ &    Noise $\Delta_P$   & \multirow{2}{*}{$\ell_\text{knee}$} & \multirow{2}{*}{$\alpha_\text{knee}$} & \multirow{2}{*}{Configuration} \\
        (GHz)        &      (arcmin)        & (${\rm \mu}$K-arcmin) &                                     &                                       &                                \\
\hline
                  20 &                   11 &                  13.6 &                                 150 &                                   2.7 &          \multirow{3}{*}{both} \\
                  30 &                   73 &                  3.53 &                                  60 &                                   1.7 &                                \\
                  40 &                   73 &                  4.46 &                                  60 &                                   1.7 &                                \\
\hline
                  85 &                   26 &                  0.88 &                                  60 &                                   1.7 &         \multirow{4}{*}{split} \\
                  95 &                   23 &                  0.78 &                                  60 &                                   1.7 &                                \\
                 145 &                   26 &                  1.23 &                                  60 &                                   3.0 &                                \\
                 155 &                   23 &                  1.34 &                                  60 &                                   3.0 &                                \\
\hline
                  90 &                   26 &                  0.42 &                                  60 &                                   1.7 &     \multirow{2}{*}{non-split} \\
                 150 &                   23 &                  0.64 &                                  60 &                                   3.0 &                                \\
\hline
                 220 &                   13 &                  3.48 &                                  60 &                                   3.0 &          \multirow{2}{*}{both} \\
                 270 &                   13 &                  5.97 &                                  60 &                                   3.0 &                                \\
\hline
\end{tabular}
\caption{Technical specifications for a CMB-S4-like Small Aperture Telescope and 20~GHz channel for a Large Aperture Telescope from~\cite{S4pipeline25}. We show the frequency band centers, FWHM apertures, polarization white-noise levels, and atmospheric noise parameters. The total noise power spectrum is modeled by \eqref{eqn:noise}. The configuration column indicates in which experimental setup the channel is being used, with `split' being the baseline configuration proposed for CMB-S4.}
\label{tab:S4specs}
\end{table}

\subsection{Sky Patch}
The map-based simulations used in this work assume a sky patch in the cleanest region of the Southern Galactic sky, similar to the footprint of the BICEP experiment \cite{BICEP2021}.
The observed sky patch is defined by a relative depth map compatible with observations from the South Pole~\cite{CMBS42022, S4pipeline25}.
We center the patch at (RA~=~$10^\circ$, dec~=~$-45^\circ$), with a Small Aperture Telescope (SAT) field of view of $29^\circ$ and boresight scans covering a rectangular patch of size ($\Delta$RA~=~$90^\circ$, $\Delta$dec~=~$5^\circ$).
The Large Magellanic Cloud (LMC), Small Magellanic Cloud (SMC), and Galactic foregrounds confine the patch on all sides.
With this patch selected, we calculate the sky fraction of the relative depth map to be $2.5\%$ using
\begin{align}
    f_\text{sky} &= \frac{1}{n_\text{pix}}\sum_{p} w_p^2
    \label{eqn:fsky}
\end{align}
where $w_p$ is the window function that describes an apodized mask shown in Figure~\ref{fig:experiment}. For the purpose of this work we assume that all instruments are observing the same patch from the same site at the South Pole.

\subsection{Noise Model}

The noise model assumed here has a white noise part and a $1/f$ noise component that captures the residual low-frequency signal after time-domain filtering. 
The noise power spectrum, $N_\ell$, is modeled for each channel as:
\begin{align}
    N_\ell = \Delta^2_P \left[1 + \left(\frac{\ell}{\ell_\text{knee}}\right)^{-\alpha_\text{knee}} \right],
    \label{eqn:noise}
\end{align}
where $\Delta_P$ is the white noise level, and $\ell_{\rm knee}$ and $\alpha_{\rm knee}$ specify the low-$\ell$ contribution from the projected $1/f$ noise. 
The values of these three noise parameters are frequency dependent.
The parameter values used in our simulation are listed in Table~\ref{tab:S4specs}, and are based on the most recently published CMB-S4 forecasting paper \cite{S4pipeline25}. 

The baseline CMB-S4 design assumes a so-called `split' band configuration which deploys two types of dichroic detectors at CMB frequencies, splitting each of the two CMB-sensitive atmospheric windows to make two separate frequency bands. 
This results in 85 and 95 GHz bands in the first atmospheric window and 145 and 155 GHz in the second. 
This design is different from stage-3 experiments like SO, which deploy only one type of dichroic detector that does not split the two atmospheric windows. 
Having more frequency channels should in principle allow us to mitigate issues of foreground decorrelation, however the splitting of the atmospheric window reduces the CMB-sensitivity of the detectors. 
To study the interplay of the design choices with foreground mitigation we also consider a configuration with `non-split' CMB bands, keeping the total number of detectors unchanged between the two configurations. 
The non-split CMB-bands at 90 and 150 GHz have improved instrumental noise and twice the number of detectors of the individual split bands. 

In this work, we model the noise level of non-split bands as the average noise level of the pair of corresponding split bands divided by two (combined sensitivity, a factor $\sqrt{2}$ better for the wider frequency band, and a factor $\sqrt{2}$ better for twice the number of detectors). 
We assign the larger of the two MF beams to the 90 GHz channel. 
The other frequencies are identical for the two configurations.
The noise parameters and resolution of both configurations are shown in Table~\ref{tab:S4specs}.

We note here that the ultra-low frequency 20 GHz channel is intended for deployment on a Large Aperture Telescope (LAT) with different field of view and low-$\ell$ noise properties than those of SAT channels. 
This frequency channel is intended to help with the synchrotron emission in the higher multipole range. 
For simplicity, we assume here that the 20 GHz on the LAT would have the same hit map as that of the SAT. 
This can be approximately achieved with a proper scan strategy.

\begin{figure}[t]
    \centering
    \includegraphics[width=0.46\linewidth]{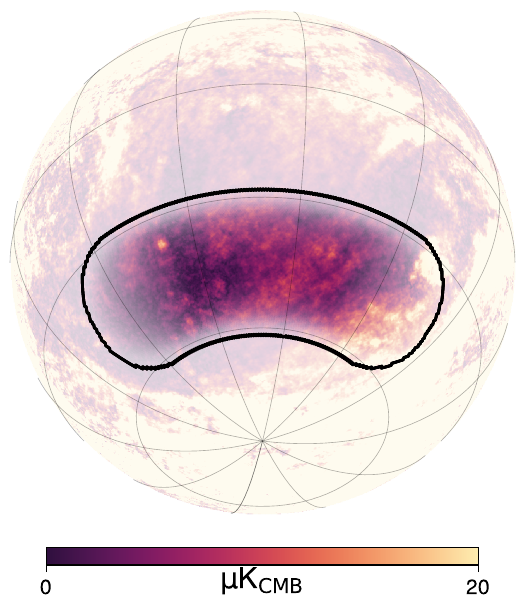}
    \includegraphics[width=0.53\linewidth]{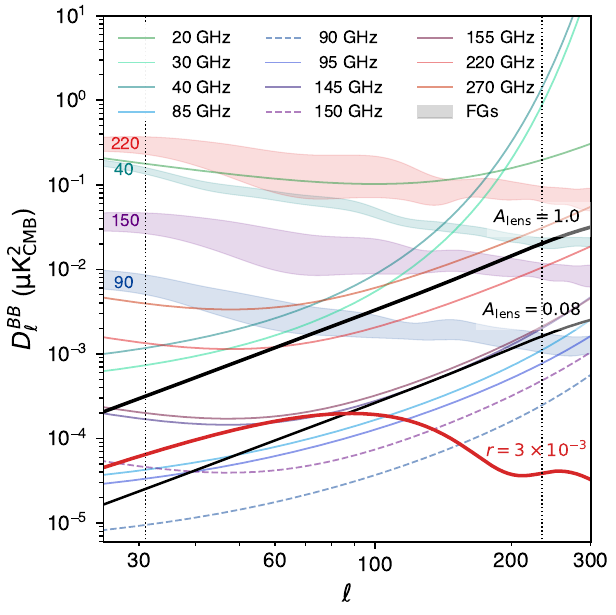}
    \caption{Sky patch (left) with binary outline traced in black, centered at (RA~=~$10^\circ$, dec~=~$-45^\circ$). The apodization yields $f_\text{sky}=2.5\%$ as calculated with Eq.~\eqref{eqn:fsky}. The sky map is polarized intensity of the PySM3 high-complexity foregrounds at 270~GHz. Plot of beam deconvolved noise curves (right) from noise levels in Table~\ref{tab:S4specs} for both experimental configurations.
    We also plot the expected foregrounds from PySM (shaded areas) at select frequencies (40, 90, 150, and 220 GHz) where the width is given by the discrepancy between the low, medium, and high complexity models. Overlaid are the theoretical CMB signals -- lensing signal (black) and primordial $B$-mode signal (red). The $\ell$-range used in the analysis is bounded by the dotted vertical black lines.}
    \label{fig:experiment}
\end{figure}

Map-based noise simulations are performed by first generating white noise maps from a normal, unit variance distribution.
We then perform a harmonic transform, apply the $\ell$-scaling from \eqref{eqn:noise} on the harmonic coefficients, then return to map domain, and finally rescale these maps by the white noise level of the channel.
The theory curves for this noise model are plotted in Figure~\ref{fig:experiment}, along with some theoretical CMB spectra to indicate the required sensitivity.
We then multiply the noise map by the relative hits map to obtain the anisotropic noise.
This process is performed for the $TEB$ modes separately. 
A final step produces $IQU$ noise maps from the $TEB$-mode noise maps by harmonic transforms.

\subsection{Sky Models}
\label{sec:skymodels}

In our simulation pipeline we use the recommended foreground model suites from the Python Sky Model 3 (PySM3)~\cite{Thorne2017, Zonca2021, PanEx25}. 
In this work we only include the polarized foreground components, ignore all extragalactic foregrounds, and use monochromatic bandpasses, so we also ignore CO line-emissions. 
The large scale modes of the dust models from PySM3 are based on the Planck GNILC dust map, while the polarized synchrotron are based on the WMAP K-band observations. 
The small scales are generated from best-fit power law model of the dust/synchrotron power spectra. 
The templates are defined at 353 GHz for the dust and 23 GHz for the synchrotron. 
The foregrounds at other frequencies are produced by rescaling these templates to those frequencies assuming a modified blackbody emission law for the dust or a power law (with or without spectral running) emission for the synchrotron. 

The low-complexity foregrounds (PySM3 synchrotron model \texttt{s4}, PySM3 dust model \texttt{d9}) uses templates produced as described above, and their frequency scaling is constant over the entire sky. 
The medium-complexity model (PySM3 synchrotron model \texttt{s5}, PySM3 dust model \texttt{d10}) uses the exact same spatial template as the low-complexity, with spatial variations in the frequency scaling.
The high-complexity foregrounds (PySM3 synchrotron model \texttt{s7}, PySM3 dust model \texttt{d12}, PySM3 anomalous microwave emission model \texttt{a2}) include an additional polarized foreground in form of the anomalous microwave emission (AME) based on the Planck Commander AME data product. 
The synchrotron model introduces a spectral running in the power law frequency scaling, while the dust is a multi-layer model with line-of-sight decorrelation generated following the prescription of \cite{2018MNRAS.476.1310M}. 
We refer the reader to tables 3, 4 and 7 of \cite{PanEx25} for a tabulated summary of the PySM3 foreground models.
These recommended PySM3 foreground suites span the current range of foreground modeling uncertainty, and the power spectra of select frequencies are plotted in Figure~\ref{fig:experiment}.
Going through the model suites from low to high, increases the number polarized foreground components and frequency decorrelation for the dust and synchrotron. 
This makes the component separation problem more difficult as we go from low to medium to high complexity foregrounds. 

\begin{figure}[t]
    \centering
    \includegraphics[width=\linewidth,trim={0 5cm 0 4cm},clip]{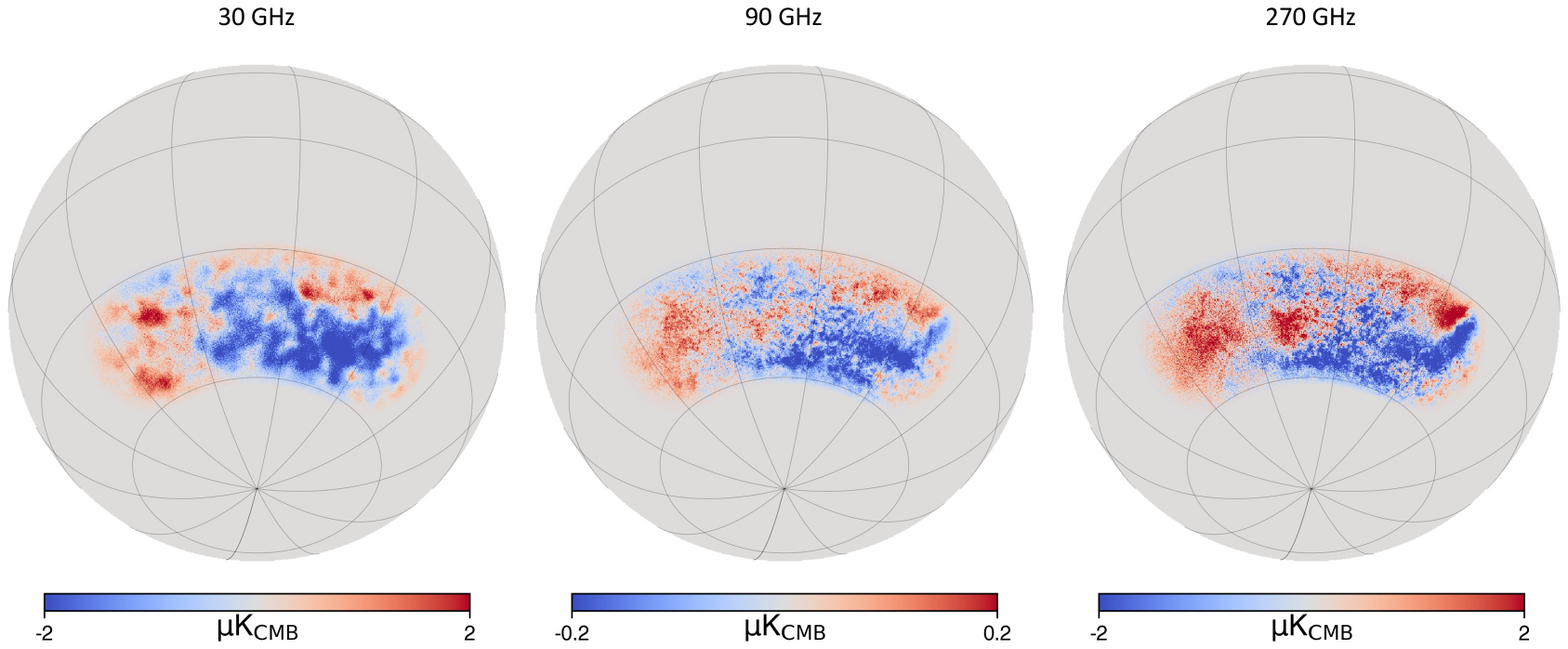}
    \caption{Maps of total simulated $B$-mode observations in three of the frequency channels, illustrating the contributions at low frequency dominated by synchrotron (left), at the highest frequency dominated by dust (right), and in a CMB channel (middle). The CMB signal is assumed to be delensed with $A_{\rm lens}=0.08$, and in this example includes tensor modes with $r=3\times10^{-3}$.}
    \label{fig:skymaps}
\end{figure}

The CMB maps are generated as Gaussian random fields with the CMB spectrum generated using \texttt{CAMB}\footnote{\url{https://camb.info}} \cite{CAMB} with Planck 2018 cosmology \cite{PL2018}.
For simplification we avoid map level delensing in this work and assume an achieved delensing of $A_\text{lens}=0.08$ based on \cite{Belkner2024}. 
We implement this by multiplying our lensed CMB $B$-mode maps by a factor of $A_\text{lens}^{1/2}$ before coadding with other maps and for the $r=3\times 10^{-3}$ case, we also add tensor mode-only CMB realizations. 
The CMB and foreground maps are coadded to produce the input sky signal and then smoothed with a Gaussian beam, with the FWHMs provided in Table~\ref{tab:S4specs}, before then being coadded with the noise maps to complete our mock observation. 
Mock observations for three of the frequency channels and for the high-complexity foregrounds are displayed in Figure~\ref{fig:skymaps}.

\section{Pipeline}
\label{sec:pipeline}

Once we have the mock observations in the form of $IQU$ maps, we can proceed with component separation and measurement of the primordial $B$ modes.
First we convert the $Q$ and $U$ polarization maps into $B$-mode maps with template-based cleaning of $E$-to-$B$ leakage~\cite{Liu19}.
The spherical harmonic coefficients of each $B$-map, corrected by the beam, are given by $\hat d_{\ell m}$, which are vectorized over frequencies as $\bm{\hat d}_{\ell m}$.
We use the healpy\footnote{\url{https://healpix.sourceforge.io}}~\cite{healpy} \texttt{anafast} function to compute the power spectrum and correct for sky fraction by dividing by $f_{\rm sky}$, and bin with weights of ($2\ell + 1$) to account for the number of modes at each $\ell$.
The resultant data covariance matrix can be written as
\begin{align}
    \bm{\mathsf{\hat{C}}}_q &= \frac{1}{n_q} \sum_{\ell\,\in\,\mathcal{D}_q}  \sum_{m}\bm{\hat d}_{\ell m}\bm{\hat d}_{\ell m}^\dagger
    \label{eqn:covmat}
\end{align}
where $\mathcal{D}_q$ is the binning of $\ell$ into $Q$ bins, each ranging over $\ell_\text{min} \leq \ell \leq \ell_\text{max}$, and the number of modes is given by $n_q = \left((\ell_\text{max} + 1)^2 - (\ell_\text{min})^2\right) f_\text{sky}$.

\subsection{SMICA}
\label{sec:smica}

The $B$-mode sky emission at millimeter wavelengths is well modeled as a linear combination of the CMB, astrophysical foregrounds, and instrumental noise. We assume that the sky signal can be decomposed in terms of $n_c$ independent components whose spatial variability is entirely captured by the spherical harmonic coefficients of templates $\bm{s}_{\ell m}$, vectorized over number of components.
The data model $\bm{d}_{\ell m}(\theta)$, vectorized over frequencies, is then written as:
\begin{align}
    \bm{d}_{\ell m}(\theta) &= \bm{\mathsf{A}} \bm{s}_{\ell m} + \bm{n}_{\ell m},
\end{align}
where $\bm{\mathsf{A}}$ is the mixing matrix whose column space defines how the independent component templates scale to different frequencies, and $\bm{n}_{\ell m}$ are the spherical harmonic coefficients from the noise maps, corrected by the beam, vectorized over frequencies.
Frequency vectors have length $n_f$, the number of frequency channels and component vectors have length $n_c$, the number of signal components in our linear superposition. 
One of the components in $n_c$ is the CMB, so we will denote the number of foreground components as $n_\text{FG} = n_c - 1$.
We use the binning $\mathcal{D}_q$ and write the SMICA data covariance model as:
\begin{align}
    \bm{\mathsf{C}}_q(\theta) &= \bm{\mathsf{A}} \bm{\mathsf{S}}_q \bm{\mathsf{A}}^\dagger + \bm{\mathsf{N}}_q.
    \label{eqn:modelcovmat}
\end{align}
Here the mixing matrix $\bm{\mathsf{A}}$ is $n_f \times n_c$, each of the signal covariance matrices $\bm{\mathsf{S}}_q$ is $n_c \times n_c$, and each of the noise covariance matrices $\bm{\mathsf{N}}_q$ is $n_f \times n_f$.
The matrices $\bm{\mathsf{A}}$ and $\bm{\mathsf{S}}_q$ are unknown. 
The noise covariance matrix is assumed to be diagonal, and is estimated from the variance of the noise auto-spectra computed from simulations.
We want to constrain the elements of the unknown matrices but we reduce the number of unknowns by making a few simplifications without any loss of generality. 
The column-space of the mixing matrix gives the frequency scaling of the different components. 
Since our data is in units of CMB temperature, the column-vector corresponding to the CMB is an all-ones vector. 
That leaves $n_{\rm FG}$ unknown mixing vectors. 
In parametric component separation methods, the frequency scaling is assumed to have some particular analytic form (e.g. modified blackbody for the dust). 
However, in our implementation of SMICA for this work we do not assume any analytic form for the frequency scaling of any of the foreground components.

Each of these mixing vectors have degeneracy of an overall scaling factor that can be exchanged with the scaling of the covariance of the component modeled in $\bm{\mathsf{S}}_q$. 
To resolve this degeneracy, we identify pivot frequencies for each of the $n_{\rm FG}$ mixing vectors where they are normalized to 1. 
The pivot frequencies for components 1 through 4 are: 270 GHz, 20 GHz, 155 GHz (150 GHz) and 30 GHz for split (non-split) configuration.
Additionally, the 1st component mixing vector is set to zero at the lowest frequency, while the 2nd component mixing vector is set to zero at the two highest frequencies due to low signal from these components in these channels. 

The signal covariance matrix for each $\ell$-bin is a cross-component covariance matrix. 
The CMB auto-term is the only analytic form assumed in this work, modeled from the $B$-mode power spectrum:
\begin{align}
    C_\ell^\text{CMB} &= r C_\ell^{\text{tens}\,r=1} + A_\text{lens} C_\ell^\text{lens}.
\end{align}
In the CMB model, the only parameter we are fitting for is $r$; the lensing residual is the same as was used to generate the CMB maps in section~\ref{sec:skymodels}, and the tensor spectrum is based on Planck 2018 cosmology just as when generating the maps. 
This spectrum is then binned with weights of ($2\ell+1$), as for the covariance matrix. 
Since the CMB is uncorrelated with the foregrounds, we set all cross-terms with the CMB to zero. 
Additionally, we assume that the third and fourth foreground components are independent and do not correlate with each other nor with the first two foreground components. 
This is justified since any correlated part of the foreground should be captured by the first two components.
We allow the first two foreground components to have non-zero cross-correlation.

Now that we have defined our data covariance model, we obtain the best-fit model parameters by sampling the SMICA log-likelihood defined as:
\begin{align}
    \ln\mathcal{L}(\theta) &= -\tfrac{1}{4}\sum_{q\,=\,1}^Q n_q D(\bm{\mathsf{\hat{C}}}_q, \bm{\mathsf{C}}_q(\theta)).
    \label{eqn:loglikelihood}
\end{align}
Here $D(\bm{\mathsf{R}}_1, \bm{\mathsf{R}}_2)$ is the Kullback-Leibler divergence between two $n \times n$ matrices
\begin{align}
    D(\bm{\mathsf{R}}_1, \bm{\mathsf{R}}_2) &= \text{tr}\left(\bm{\mathsf{R}}_1 {\bm{\mathsf{R}}_2}^{-1}\right) - \ln\det\left(\bm{\mathsf{R}}_1 {\bm{\mathsf{R}}_2}^{-1}\right) - n
    \label{eqn:divergence}
\end{align}

\subsection{Implementation}
\label{sec:implementation}

\begin{figure}[t]
    \centering
    \includegraphics[width=\linewidth,trim={0 1.2cm 0 2cm},clip]{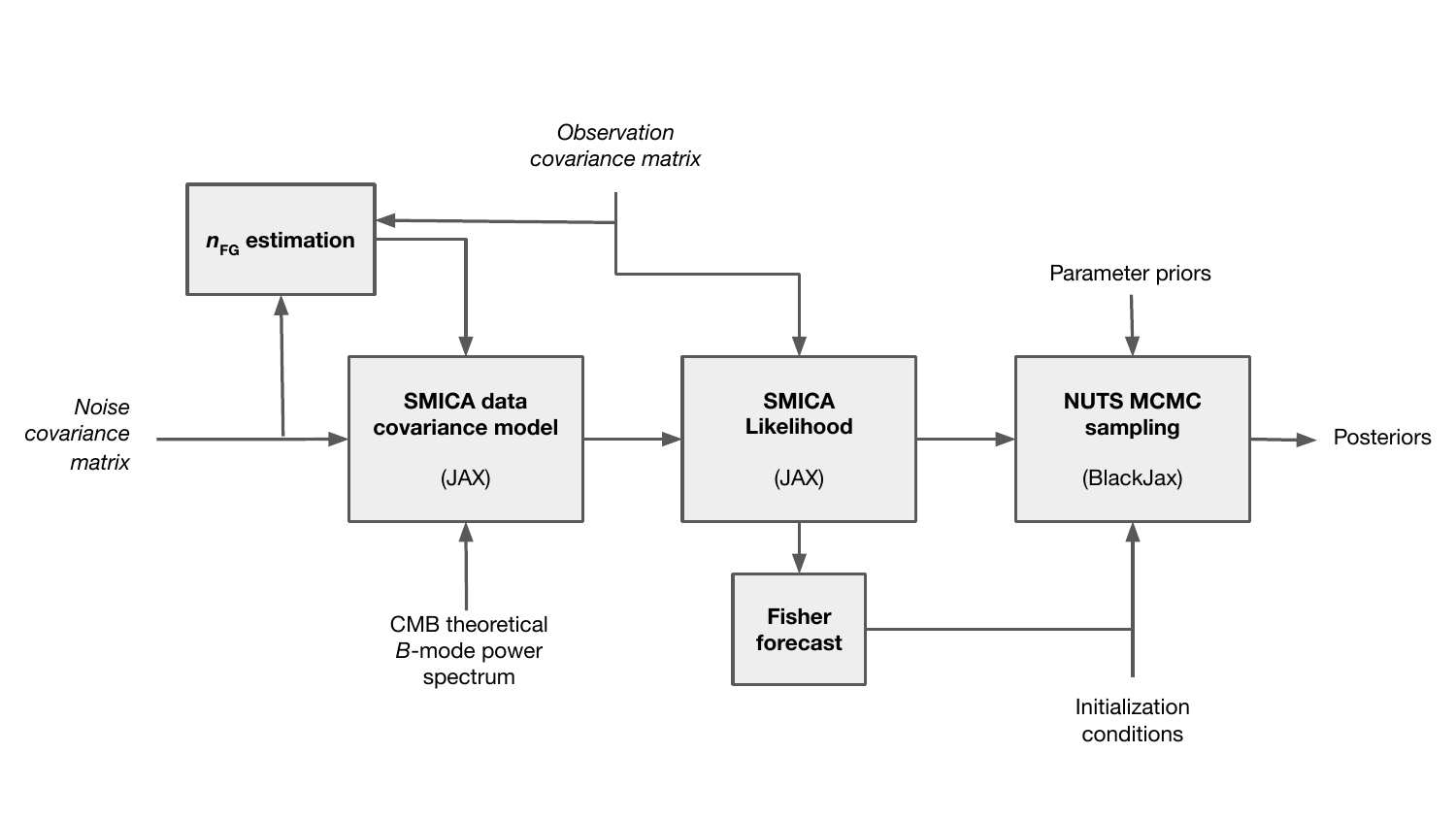}
    \caption{Flowchart describing our SMICA pipeline. The boxes indicate the main parts of the SMICA pipeline. The covariance matrices are computed as in Eq. \eqref{eqn:covmat} and are the main inputs to the pipeline.}
    \label{fig:flowchart}
\end{figure}

The SMICA data covariance model of Eq. \eqref{eqn:covmat}, (negative) log-likelihood of Eq. \eqref{eqn:loglikelihood}, and the sampling setup form the core of the SMICA pipeline. 
A flowchart for this pipeline is shown in Figure \ref{fig:flowchart}.
The covariance model and likelihood are implemented in \texttt{JAX}~\cite{JAX}, making use of \texttt{JAX}'s just-in-time compilation and vectorized-mapping features, allowing for fast computations. 
The log-likelihood is sampled with a No-U-Turn Sampler (NUTS) which pairs gradient descent with Monte-Carlo (MC) sampling. 
The NUTS sampling is implemented with BlackJax \cite{BlackJax}, and makes use of JAX's automatic differentiation features for computing the gradients of the log-likelihood. 
The automatic differentiation also provides fast Fisher forecasts, which we use to inform our initial conditions. 
The choice of $n_{\rm FG}$ for the model is informed from the noise-whitened observation covariance matrix similar to~\cite{Remazeilles2011}, and is described in Sec \ref{sec:svd}.

The likelihood calculation involves computing the inverse of the model covariance matrix which is not numerically stable for all combinations of parameter values. 
To aid the stability we rescale the data by a global scaling factor since the log-likelihood is invariant under such scaling. 
This scaling factor is appropriately chosen to ensure that the covariance model determinant is close to unity.

To initialize the MC, we perform a singular value decomposition (SVD) of the foreground covariance matrix.
We recognize this cannot be done on a real sky measurement.
However, one can perform a SVD on the full data covariance matrix or on the `noise-whitened' data covariance matrix as is done in the generalized needlet internal linear combination method~\cite{Remazeilles2011}.
We verified on a single realization of non-split, high-complexity foregrounds that using the full data covariance matrix gave similar results, though this only works in the regime of CMB signal being subdominant compared to foreground components that are being fit for.
The singular vectors for the lowest-$\ell$ bin is used to initialize the elements of the mixing matrix.
We then perform SVDs in each bin to determine the singular values of each vector.
These singular values are the power spectra as a function of $q$.
We use the Fisher forecast of the parameter uncertainties to add random fluctuations to the initializations.
The cross spectrum between synchrotron and dust is initialized at zero. 

While the NUTS sampler does not require priors, we do set some priors in our model to prevent degeneracies and unphysical regimes.
We enforce a positive prior on the mixing vectors. 
Indeed, frequency scalings for Galactic foregrounds are not expected to change sign as a function of frequency. 
Relaxing this condition, however, does not change the results presented here.
Additionally, we enforce a positive prior on the power spectrum of the CMB and the first two foreground components.
This means that $r$ can be negative, though not so negative that subtracting tensor power from the delensed scalar power gives a negative difference.
This also allows for a negative power spectrum in the third and fourth components of the foreground signal -- negative values here are not unphysical because these are corrections on dust and synchrotron, not additional astrophysical foregrounds.

\subsection{Goodness of Fit}

The generality of the SMICA model means it is constructed with a large number of parameters, which we will call $n_\text{params}$.
The number of parameters for the $n_\text{FG}=2$ model is 29 and 25 for split and non-split configurations respectively.
In the case of $n_\text{FG}=3$, the number of parameters increases to 42 and 36, and for $n_\text{FG}=4$ this rises to 55 and 47 parameters in the two configurations. 
These numbers are reduced from the fully general case because some elements of the mixing and signal matrices are fixed to one or zero in our construction of the model, as explained in section~\ref{sec:smica}.
To justify the increase in $n_\text{params}$ when incorporating additional components, we need to quantify the goodness of fit, which we define as
\begin{align}
    \chi^2/n_\text{dof} &= \frac{-2\ln\mathcal{L}(\theta_\text{MAP})}{\tfrac{1}{2} \, n_f (n_f+1) Q - n_\text{params}}.
\end{align}
Here $\theta_\text{MAP}$ represents the parameters at the maximum a-posteriori (MAP), which gives the best fit parameters for the SMICA model.

\section{Results}
\label{sec:results}

\begin{figure}[tb]
    \centering
    \includegraphics[width=0.48\linewidth]{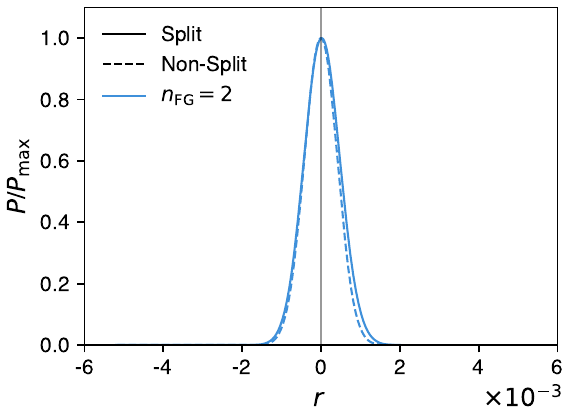}
    \includegraphics[width=0.48\linewidth]{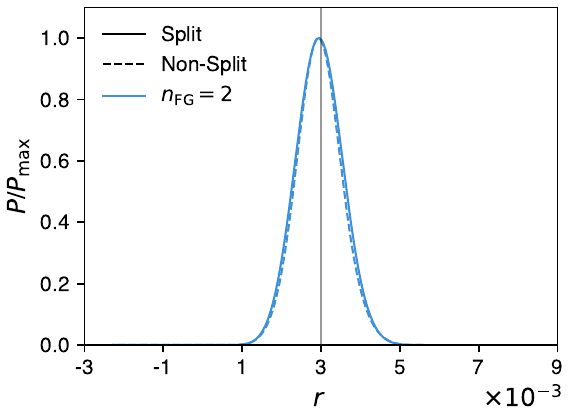}
    \caption{SMICA posterior of $r$ in the case of low-complexity foregrounds for both $r=0$ (left) and $r=3\times10^{-3}$ (right), with measurement values are given in Tables~\ref{tab:r=0},~\ref{tab:r=0.003}. Both experimental configurations are shown, and only the $n_\text{FG}=2$ model was used. The theory value of $r$ used for the input maps is marked by the vertical gray line.}
    \label{fig:rLow}
\end{figure}

We validate our pipeline with the low-complexity foregrounds which have only two rigid-scaling emission components; therefore we only need two foreground components to obtain an unbiased fit for $r$. 
Figure~\ref{fig:rLow} shows the posterior probability distribution (normalized at its peak). 
Tables \ref{tab:r=0} and \ref{tab:r=0.003} show the MAP estimate of $r$, the uncertainty, $\sigma_r$, (computed as the standard deviation of the MC chains) and the $\chi^2/n_\text{dof}$ goodness-of-fit metric for $r=0$ and $r=3\times 10^{-3}$. 
Note that the expectation value for $r$ (minimum mean squared error estimate) is nearly identical to the MAP estimate. 
We obtain an unbiased fit for $r$ for both instrument configurations and the standard deviation $\sigma_r$ is small, exceeding the CMB-S4 $r$-sensitivity target. 
Additionally, the standard deviation obtained matches the Fisher forecast for $\sigma_r$ which validates our sampling method. 
The $\chi^2/n_\text{dof}$ metric is close to one, indicating a reasonably good fit.
When we attempt this fit to low-complexity foregrounds with $n_\text{FG} > 2$, we cause a degeneracy in the foreground parameters.
When we fit for 3 or 4 foreground components instead of 2, instabilities are observed in the MC chains for the foreground parameters specifically.
We also find the $\chi^2/n_\text{dof}$ increases because we are adding parameters to the model without obtaining a better fit to the data.
Though we have degeneracies and lack of convergence in the foregrounds, the posterior of the $r$ parameter specifically remains well constrained and remains largely unchanged, showing that the method is robust.

\begin{figure}[t]
    \centering
    \includegraphics[width=0.48\linewidth]{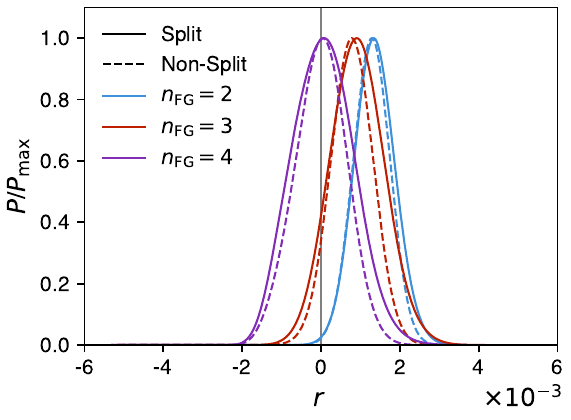}
    \includegraphics[width=0.48\linewidth]{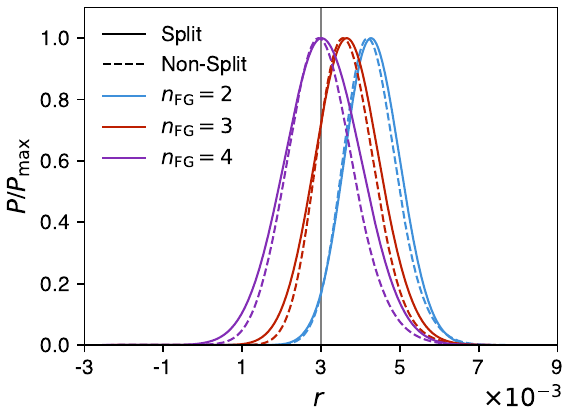}
    \caption{SMICA posterior of $r$ in the case of medium-complexity foregrounds for both $r=0$ (left) and $r=3\times10^{-3}$ (right), with measurement values are given in Tables~\ref{tab:r=0},~\ref{tab:r=0.003}. Both experimental configurations are shown, with the number of foreground components ($n_\text{FG}$) used in the SMICA model marked by different colors. The theory value of $r$ used for the input maps is marked by the vertical gray line.\\}
    \label{fig:rMedium}
    \centering
    \includegraphics[width=0.48\linewidth]{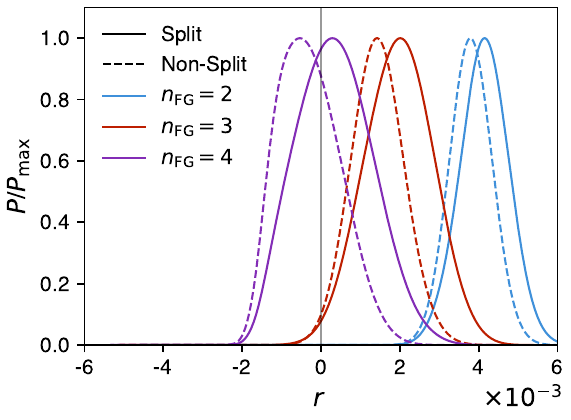}
    \includegraphics[width=0.48\linewidth]{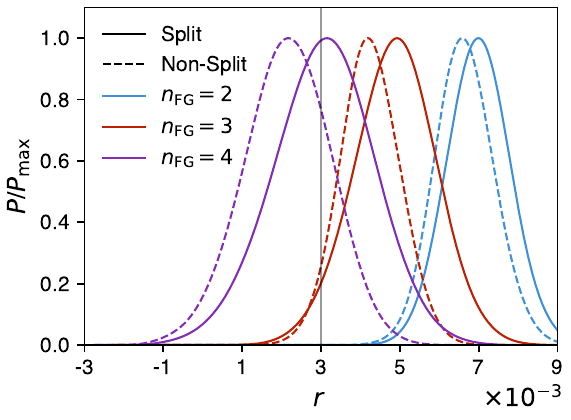}
    \caption{SMICA posterior of $r$ in the case of high-complexity foregrounds for both $r=0$ (left) and $r=3\times10^{-3}$ (right), with measurement values are given in Tables~\ref{tab:r=0},~\ref{tab:r=0.003}. Both experimental configurations are shown, with the number of foreground components ($n_\text{FG}$) used in the SMICA model marked by different colors. The theory value of $r$ used for the input maps is marked by the vertical gray line.}
    \label{fig:rHigh}
\end{figure}

However, for both the medium- and high-complexity foregrounds, our MAP estimate of $r$ with a two component model are inconsistent with the inputs, showing a clear bias as demonstrated in Figures~\ref{fig:rMedium} and~\ref{fig:rHigh}.
In general we find higher bias and uncertainty for the high-complexity foreground. 
One mixing vector for dust and one for synchrotron are clearly not enough to fully model the foregrounds; the spatial variation of the dust/synchrotron frequency scaling or the line-of-sight decorrelation (high-complexity dust) require additional independent components to model them in the SMICA model. 
When we do not have additional components to model the added complexity, part of the unmodeled foreground power is projected onto the CMB vector.

To remove the bias from our $r$-measurement, we introduce additional independent components to model the residual foreground power that is not modeled by the first two components.
For this work, we introduce either one or two additional components, taking $n_\text{FG}$ from two to three or four.
The need for such additional components, and how many are needed, is discussed further in appendix~\ref{sec:svd}.
We find that we need four components to adequately model either the medium- or high-complexity foregrounds so that we make an unbiased estimate of $r$, as is shown in Figures~\ref{fig:rMedium} and~\ref{fig:rHigh}. 
Increasing the number of independent components increases the number of parameters in the model. 
This increases the standard deviation, as shown in Tables~\ref{tab:r=0} and~\ref{tab:r=0.003}, demonstrating the bias-variance tradeoff for SMICA. 
We also note that for the $r=0$ case the obtained posteriors are slightly skewed, owing to the positivity prior on the total CMB power spectrum model. 

While $\sigma_r$ increases with the number of parameters, the $\chi^2/n_\text{dof}$ generally improves with additional components in our model as is indicated by Tables~\ref{tab:r=0} and~\ref{tab:r=0.003} up to $n_\text{FG}=4$ for both medium and high complexity foregrounds.
It is important to note that $\chi^2/n_\text{dof}$ is not far from one anywhere, even in the cases with significant $r$-bias. 
This is most likely because the relative contribution of the primordial $B$-mode to the data covariance model is vanishingly small. 
The foreground residuals in the CMB frequencies are consistent with zero within $1\sigma$ in all models and configurations, as shown in appendix \ref{sec:foregrounds}.  
This implies that the $\chi^2/n_\text{dof}$ is not very useful when it comes to informing us about the $r$-bias.
We do not find any $\chi^2/n_\text{dof}$ metric smaller than 1, so even with the additional parameters there is no indication that we are over-fitting the system.

\begin{table}[tb]
\centering
\begin{tabular}{c|c|c|c|c|c|c|c}
\multirow{3}{*}{} & \multirow{3}{*}{$n_\text{FG}$} & \multicolumn{2}{c|}{Low  Complexity} & \multicolumn{2}{c|}{Medium Complexity} & \multicolumn{2}{c}{High Complexity} \\
\cline{3-8}
&& $r_\text{MAP}\pm\sigma_r$ & \multirow{2}{*}{$\chi^2/n_\text{dof}$} 
 & $r_\text{MAP}\pm\sigma_r$ & \multirow{2}{*}{$\chi^2/n_\text{dof}$}
 & $r_\text{MAP}\pm\sigma_r$ & \multirow{2}{*}{$\chi^2/n_\text{dof}$}\\     
                           &     &   ($10^{-3}$)  &      &    ($10^{-3}$)   &      &    ($10^{-3}$)  &      \\
\hline
\multirow{3}{*}{Split}  &  2  &  $0.0 \pm 0.3$ & 1.27 &   $1.3 \pm 0.4$  & 1.16 &  $4.1 \pm 0.5$  & 1.37 \\
                           &  3  &       -        &   -  &   $0.9 \pm 0.6$  & 1.17 &  $2.0 \pm 0.8$  & 1.27 \\
                           &  4  &       -        &   -  &   $0.1 \pm 0.7$  & 1.16 &  $0.3 \pm 0.9$  & 1.19 \\
\hline
\multirow{3}{*}{Non-Split} &  2  &  $0.0 \pm 0.3$ & 1.19 &   $1.3 \pm 0.3$  & 1.24 &  $3.8 \pm 0.4$  & 1.73 \\
                           &  3  &       -        &   -  &   $0.8 \pm 0.4$  & 1.16 &  $1.4 \pm 0.6$  & 1.33 \\
                           &  4  &       -        &   -  &   $0.0 \pm 0.6$  & 1.10 & $-0.5 \pm 0.7$  & 1.26 \\
\hline
\end{tabular}
\caption{SMICA measurements of $r$ in the $r=0$ case including goodness of fit ($\chi^2/n_\text{dof}$). We show results for both experimental configurations, different foreground complexities, and varying number of foreground components ($n_\text{FG}$) used in the SMICA model.\\}
\label{tab:r=0}
\centering
\begin{tabular}{c|c|c|c|c|c|c|c}
\multirow{3}{*}{} & \multirow{3}{*}{$n_\text{FG}$} & \multicolumn{2}{c|}{Low Complexity} & \multicolumn{2}{c|}{Medium Complexity} & \multicolumn{2}{c}{High Complexity} \\
\cline{3-8}
&& $r_\text{MAP}\pm\sigma_r$ & \multirow{2}{*}{$\chi^2/n_\text{dof}$} 
 & $r_\text{MAP}\pm\sigma_r$ & \multirow{2}{*}{$\chi^2/n_\text{dof}$}
 & $r_\text{MAP}\pm\sigma_r$ & \multirow{2}{*}{$\chi^2/n_\text{dof}$}\\     
                           &     &   ($10^{-3}$)  &      &    ($10^{-3}$)   &      &    ($10^{-3}$)  &     \\
\hline
\multirow{3}{*}{Split}  &  2  &  $2.9 \pm 0.5$ & 1.33 &   $4.3 \pm 0.6$  & 1.16 &   $7.0 \pm 0.7$ & 1.28 \\
                           &  3  &       -        &   -  &   $3.7 \pm 0.8$  & 1.20 &   $4.9 \pm 1.0$ & 1.26 \\
                           &  4  &       -        &   -  &   $3.0 \pm 0.9$  & 1.17 &   $3.1 \pm 1.2$ & 1.21 \\
\hline
\multirow{3}{*}{Non-Split} &  2  &  $2.9 \pm 0.5$ & 1.30 &   $4.2 \pm 0.6$  & 1.32 &   $6.6 \pm 0.7$ & 1.66 \\
                           &  3  &       -        &   -  &   $3.6 \pm 0.7$  & 1.26 &   $4.2 \pm 0.7$ & 1.34 \\
                           &  4  &       -        &   -  &   $2.9 \pm 0.8$  & 1.21 &   $2.2 \pm 1.1$ & 1.26 \\
\hline
\end{tabular}
\caption{SMICA measurements of $r$ in the $r=3\times10^{-3}$ case including goodness of fit ($\chi^2/n_\text{dof}$). We show results for different foreground complexities, both experimental configurations, and varying number of foreground components ($n_\text{FG}$) used in the SMICA model.}
\label{tab:r=0.003}
\end{table}

Across all our $r$-measurements, the increase in sensitivity of the non-split band configuration improves both the $r$-bias and $\sigma_r$, with the exception of the bias for high-complexity foregrounds modeled with 4 components. 
More frequency channels that in turn have greater noise (in the split-band configuration) do not help to reduce the bias when the residual foreground power contributing to the $r$-bias has a signal-to-noise on the order of one. 
Increasing the sensitivity of the CMB bands is more useful in fitting this residual power.
Interestingly, for the non-split band configuration, we find a slight negative bias for the high-complexity model. 
This is likely driven by the overestimate of foreground power in the first bin, as shown in Appendix \ref{sec:foregrounds}. 
The split band configuration sees a slight positive bias in this case, so we do not believe this to be a problem with the method.
Since the bias depends solely on the accuracy of the spectral fit for foreground emission, the single high-complexity foreground realization used in all 96 observations may be the source of the apparent negative bias.

\section{Discussion}
\label{sec:discussion}

There are other component separation methods which have successfully tackled the problem of $r$-bias for the medium- and high-complexity foreground models. 
However, all such methods depend on making some assumptions on foreground components. 
These work well for sky models when we know what is driving the $r$-bias \cite{Liu25}, or when the additional assumptions match what is built into the sky models used in the simulations. 
For the actual sky we lack the knowledge of the foregrounds at the sensitivities required for current and future experiment. 
SMICA provides a clear framework to model the data covariance with added components to reduce the $r$-bias without needing to make any strong assumptions on the foregrounds.

The cost of SMICA's generality is that it has many more parameters than other parametric methods, but SMICA can also be specialized to have similar parameterization as the commonly used parametric models (i.e. assuming functional forms). 
One possible path forward is a hybrid parameterization where functional forms are assumed for some parts of the full SMICA covariance model. 
This would allow us the flexibility of SMICA's free parameterization depending on the component. 

We note that as is the case for any fitting procedure, SMICA can only work if the model that is fit is not wildly underconstrained, leading degeneracies in the parameters of interest. Still, we find convergence to the correct value of $r$ even in cases for which only the foreground parameters suffer from degeneracies, as is the case when we try to fit for 3 or 4 foreground components in the low-complexity case.

\section{Conclusions}
\label{sec:conclusions}

In this work we have demonstrated that it is possible, with SMICA, to obtain unbiased estimates of $r$ without making any assumptions on the foreground properties. 
Reducing the bias comes at the cost of increased uncertainty. 
In SMICA this is achieved by modeling the residual foreground power with additional independent components. 
We have motivated how the SVD of the data covariance can inform the need for such additional components in the model. 
One important finding from this work is that the $\chi^2/n_{\rm dof}$ goodness-of-fit metric is uninformative when it comes to the $r$-bias. 
However, we still find it useful to show improvement of the fit with the added components.

\acknowledgments
Some of the results in this paper have been derived using the healpy and HEALPix packages.
Funding for this work was provided by the CMB-S4 project, and the work was conducted at Lawrence Berkeley National Laboratory.
This research used resources of the National Energy Research Scientific Computing Center (NERSC), a Department of Energy User Facility using NERSC award HEP-ERCAP0034243.

\bibliographystyle{JHEP}
\bibliography{biblio.bib}

\appendix
\section{Number of Independent Components}
\label{sec:svd}

\begin{figure}[b]
    \centering
    \includegraphics[width=\linewidth]{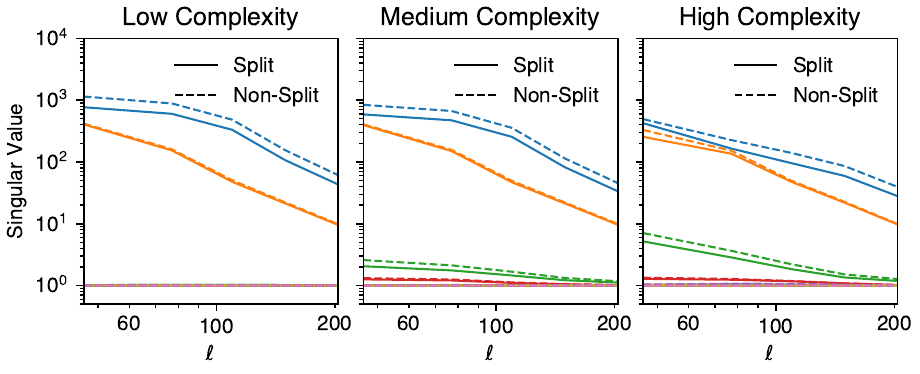}
    \caption{Noise whitened SVD singular values, plotted for each $\ell$-bin, for low-complexity (left), medium-complexity (middle), and high-complexity (right) PySM3 foregrounds from mock observation. 
    The colors indicate the different components, which we are blind to when fitting.}
    \label{fig:SVD}
\end{figure}

We can characterize our foregrounds exactly by performing an SVD of the foreground-only covariance matrix.
This cannot be done on real data, where one would be required to use the full covariance matrix to perform this same analysis.
While we use an SVD to initialize our simulation as is outlined in section~\ref{sec:implementation}, the main use of the an SVD in our pipeline is as a diagnostic tool to determine the number of foreground components needed in the SMICA model.
By adding the foreground and noise covariance matrices, noise whitening (identical to the procedure in \cite{Remazeilles2011}), and then performing the SVD, we can determine how many independent vectors are above noise level in the foregrounds.
The results of this analysis are shown in Figure~\ref{fig:SVD}, a signal-to-noise plot of every component in the foregrounds as a function of $\ell$.
With low-complexity foregrounds, we see that only two components are above significantly noise level, as we would expect.
For medium-complexity foregrounds we see four significant components, hence why we fit for up to four components in our SMICA model.
The high-complexity foregrounds also have four significant components, with greater power in the secondary components -- particularly at low-$\ell$.
These secondary components remain subdominant when compared to the primary components, however they are of the same order as the CMB signal we are trying to measure.

The power present in these subdominant components explains the source of the bias when only two components are fit; this power is mistakenly measured to be an $r$ signal, when it is just unmodeled foreground complexity.
We also see that the signal-to-noise is higher in the case of non-split bands, particularly in the blue component (which we associate most closely with dust), which is likely the source of smaller $\sigma_r$ in the non-split configuration.

We note that the $\chi^2/n_\text{dof}$ is not a great indicator of $r$-bias. 
Foreground residuals dominate the $\chi^2$, especially at low-$\ell$ where the foregrounds have orders of magnitude more power than any possible tensor modes, as we show in Appendix~\ref{sec:foregrounds}.
This is why it is critical for us to take the noise whitened SVD; we need to make an informed choice as to how many independent components to fit for in our model.
In general one would follow the prescription from \cite{Remazeilles2011} to determine the number of parameters for real data.

\section{Foregrounds}
\label{sec:foregrounds}

\begin{figure}[t]
    \centering
    \includegraphics[width=0.49\linewidth]{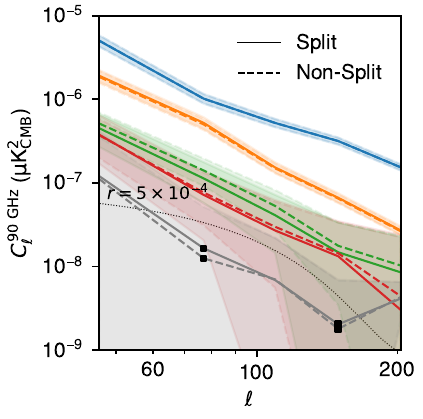}
    \includegraphics[width=0.49\linewidth]{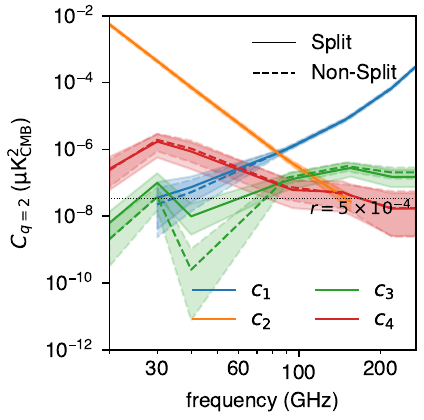}
    \caption{Plots of components in $\bm{\mathsf{A}} \bm{\mathsf{S}}_q \bm{\mathsf{A}}^\dagger$ as a function of $\ell$ (left) at 90~GHz and as a function of frequency (right) for $q=2$ (centered at $\ell=77$) as fitted by the SMICA model. Plotted with one standard deviation shaded regions for both split (solid) and non-split (dashed) configurations. The colors indicate the different components -- which we are blind to when fitting -- and match to the same components as in Figure~\ref{fig:SVD}. Additionally, the absolute value of the cross spectrum (gray) of the primary components (blue and orange) has been plotted, with the black squares indicating a negative value. Due to some values being fixed to zero in the model, the $c_1$ and $c_2$ vectors do not span the full frequency range. A theoretical $r$-signal of $5\times10^{-4}$ (dotted black) is overlaid on both plots to compare with the foreground signals.}
    \label{fig:Vecs&Specs}
\end{figure}

In our results, we focus on the posterior of $r$ because this is the measurement of a possible primordial gravitational wave signal.
However, our SMICA pipeline does not just measure $r$, it measures all the foreground components as well, including their distributions and thereby their uncertainties.
One caveat is that in SMICA we model foregrounds as independent components, so it is not straightforward to identify a component as associated with a particular type of astrophysical emission (dust/synchrotron).
This is especially true for the third and fourth foreground components that act as 'catch-all' for the residual foreground power.
In Figure~\ref{fig:Vecs&Specs} we show the best fits in the case of high-complexity foregrounds for both experimental configurations.
We see that some terms in the $\bm{\mathsf{S}}_q$ matrices, particularly in the high-$\ell$ bins, are consistent with zero within error bars. This implies that it is difficult to constrain the third and fourth foreground components, as well as the cross spectrum when they are at the noise-threshold.

\begin{figure}[t]
    \centering
    \includegraphics[width=0.48\linewidth]{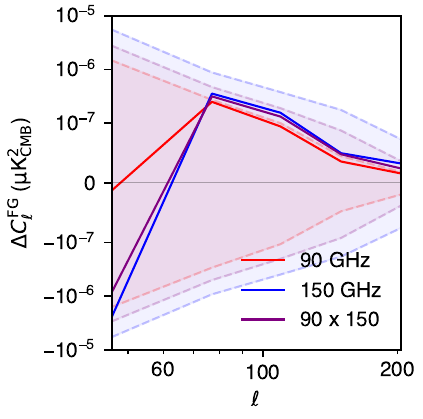}
    \includegraphics[width=0.48\linewidth]{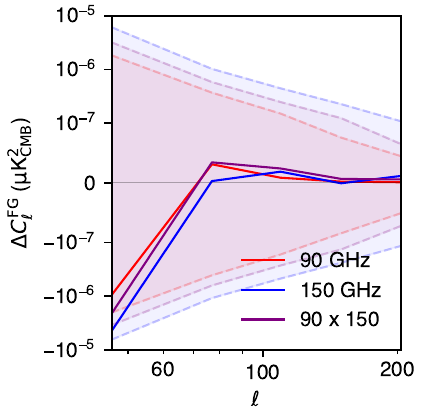}
    \caption{Foreground residuals from the high complexity, non-split, SMICA best fit plotted with $1\sigma$ shaded regions for $n_\text{FG}=2$ (left) and $n_\text{FG}=4$ (right). We define the residual foregrounds as the difference between the foreground covariance in the observation and the foreground covariance model of the SMICA best-fit, or $\Delta C_\ell^\text{FG} = \hat{C}_\ell^\text{FG} - C_\ell^\text{FG}|_\text{SMICA best-fit}$.}
    \label{fig:FGresiduals}
\end{figure}

Based on the best fit we immediately see the value of the SMICA approach; the power spectra do not follow simple power laws, and the foregrounds have complexities not fit for by simple electromagnetic laws.
Making such approximations may reduce the number of parameters, but it clearly omits details that are necessary to account for when making a measurement as precise as this on $B$ modes.

The reason why the sub-dominant foreground components are projected onto the CMB vector is clear from their frequency scaling, which is flatter than their dominant counterparts.
Given that the CMB vector is flat, the power from these sub-dominant components leaks into the CMB fit.
Because they are just above the noise level, they fall within the $1\sigma$ uncertainty of the two-component fit, as is shown in Figure~\ref{fig:FGresiduals}. 
We see that the foreground residuals are reduced by going from two to four components in the model, though in each case the foregrounds are accurate within one standard deviation.
On the right we also see the cause of the slight negative measurement of $r$, where most of the foreground residuals are close to zero with the exception of the lowest $\ell$-bin.

Since the $r$-sensitivity we wish to achieve is well below the noise level, in the high-complexity foreground case, we see a large bias in $r$ for a two-component SMICA foreground model even when the foreground residual is within one standard deviation.
When we move to four components in our model, there is less foreground residual (though with greater uncertainty) and we see no bias in $r$.
We plot just the CMB sensitive channels for high-complexity foregrounds and non-split bands, but this holds true across configurations and frequencies.

\end{document}